\documentclass[preprint,longbibliography,superscriptaddress]{revtex4-1}
\usepackage{graphicx}
\usepackage{array}
\usepackage{amssymb}
\usepackage{amsfonts}
\usepackage{amsmath}
\usepackage{color}
\usepackage{booktabs}
\usepackage{threeparttable}
\usepackage{multirow}
\usepackage{subfigure}
\usepackage{epsfig}
\usepackage{threeparttable}
\usepackage{chngpage}
\usepackage{float}
\usepackage{color}
\usepackage{bm}
\usepackage{dcolumn}
\usepackage{textcomp}
\usepackage{verbatim}
\usepackage{epstopdf}
\usepackage{multirow}
\usepackage{tabularx}
\usepackage[labelfont=bf]{caption}
\DeclareCaptionLabelSeparator{vline}{ $|$ }
\usepackage{caption}
\makeatletter
\renewcommand\@biblabel[1]{#1.}
\makeatother

\begin{document}
\draft

\title{Universal structural estimator and dynamics approximator for
complex networks}

\author{Yu-Zhong Chen}
\affiliation{School of Electrical, Computer and Energy Engineering,
Arizona State University, Tempe, Arizona 85287, USA}

\author{Ying-Cheng Lai} \email{Ying-Cheng.Lai@asu.edu}
\affiliation{School of Electrical, Computer and Energy Engineering,
Arizona State University, Tempe, Arizona 85287, USA}
\affiliation{Department of Physics,
Arizona State University, Tempe, Arizona 85287, USA}


\begin{abstract}

Revealing the structure and dynamics of complex networked systems from observed data is of fundamental importance to science, engineering, and society. Is it possible to develop a universal, completely data driven framework to decipher the network structure and different types of dynamical processes on complex networks, regardless of their details? We develop a Markov network based model, sparse dynamical Boltzmann machine (SDBM), as a universal network structural estimator and dynamics approximator. The SDBM attains its topology according to that of the original system and is capable of simulating the original dynamical process. We develop a fully automated method based on compressive sensing and machine learning to find the SDBM. We demonstrate, for a large variety of representative dynamical processes on model and real world complex networks, that the equivalent SDBM can recover the network structure of the original system and predicts its dynamical behavior with high precision.

\end{abstract}

\pacs{02.50.Le,89.75.Hc,87.23.Ge}

\maketitle

\section{Introduction} \label{sec:intro}

A central issue in complexity science and engineering is systems
identification and dynamical behavior prediction based on experimental or
observational data. For a complex networked system, often the network
structure and the nodal dynamical processes are unknown but only time
series measured from various nodes in the network can be obtained. The
challenging task is to infer the detailed network topology and the nodal
dynamical systems from the available data. This line of pursuit started
in biomedical science for problems such as identification of gene regulatory
networks from expression data in systems biology~\cite{GDA:2002,GAS:2002,
PG:2003,GBLC:2003} and uncovering various functional networks in the
human brain from activation data in neuroscience~\cite{BBAB:2007,GTF:2007,
SMRMG:2009,HLTSG:2009}. The inverse problem has also been an area of
research in statistical physics where, for example, the inverse Ising
problem in static~\cite{T:1998,CM:2011,AE:2012,NB:2012,R:2012} and
kinetic~\cite{RH:2011,MJ:2011,SBD:2011,Z:2012,ZAAHR:2013,BLHSB:2013,
DR:2014} situations has attracted continuous interest. Recent years
have witnessed the emergence and growth of a subfield of research in
complex networks: data based network identification (or reverse
engineering of complex networks)~\cite{T:2009,BL:2009,ST:2009,
CMN:2009,LP:2011,WYLKH:2011,WLGY:2011,SWFDL:2014,HSWD:2015}. In these works,
the success of mapping out the entire network structure and
estimating the nodal dynamical equations partly relies on taking advantage
of the particular properties of the system dynamics in terms of the
specific types and rules. For example, depending on the detailed
dynamical processes such as continuous-time oscillations~\cite{WCHLH:2009,
RWLL:2010,WYLKH:2011}, evolutionary games~\cite{WLGY:2011}, or epidemic
spreading~\cite{SWFDL:2014}, appropriate mathematical frameworks uniquely
tailored at the specific underlying dynamical process can be formulated
to solve the inverse problem.

In this paper, we address the following challenging question: is
it possible to develop a universal and completely data-driven framework
for extracting network topology and identifying the dynamical processes,
without the need to know a priori the specific types of network
dynamics? An answer to this question would be of significant value not
only to complexity science and engineering but also to modern data
science where the goal is to unearth the hidden structural information
and to predict the future evolution of the system. Here we introduce the
concept of {\em universal structural estimator and dynamics approximator}
for complex networked systems and demonstrate that such a framework or
``machine'' can indeed be developed for a large number of distinct types
of network dynamical processes. Our approach will be a combination of
numerical computation and physical reasoning. Since we are yet able to
develop a rigorous mathematical framework, the present work should be 
regarded as an initial attempt towards the development of a universal 
framework for network reconstruction and dynamics prediction.

The key principle underlying our work is the following. In spite of
the difference among the types of dynamics in terms of, e.g.,
the interaction patterns and state updating rules, there are two common
features shared by many dynamical processes on complex networks:
(1) they are stochastic, first-order Markovian processes, i.e., only the
current state of the system determines its immediate future; and
(2) the nodal interactions are local, i.e., a node typically interacts
with its neighboring nodes, not all the nodes in the network. The two 
features are characteristic of a \emph{Markov network} (or a \emph{Markov 
random field})~\cite{bishop:2006book,RN:2009book}. In particular, a 
Markov network is an undirected and weighted probabilistic graphical 
model that is effective at determining the complex probabilistic 
interdependencies in situations where the directionality in the 
interaction between connected nodes cannot
be naturally assigned, in contrast to the directed Bayesian
networks~\cite{bishop:2006book,RN:2009book}. A Markov network has two types 
of parameters: a nodal bias parameter that controls its preference of the
state choice, and a weight parameter characterizing the interaction strength
of each undirected link. For a network of $N$ nodes with $x_j$ being
the state of node $j$ ($j=1,\ldots,N$), the joint probability distribution 
of the state variables $\mathbf{X} = (x_1, x_2, \ldots, x_N)^{\mathrm{T}}$
is given by $P(\mathbf{X}) = \prod_{C}\phi(\mathbf{X}_C)/\sum_{\mathbf{X}}
\prod_{C}\phi(\mathbf{X}_C)$, where $\phi(\mathbf{X}_C)$ is the potential
function for a well-connected network clique $C$, and the summation in the
denominator is over all possible system state $\mathbf{X}$. If this joint
probability distribution is available, all conditional probability
interdependencies can be obtained. The way to define a clique and to
determine its potential function plays a key role in the Markov network's
representation power of modeling the interdependencies within a particular
system. 

To be concrete, in this work we pursue the possibility of modeling
the conditional probability interdependencies of a variety of dynamical
processes on complex networked systems via a binary Ising Markov network
with its potential function in the form of the Boltzmann factor, $\exp(-E)$,
where $E$ is the energy determined by the local states and their
interactions along with the network parameters (the link weights and node
biases) in a \emph{log-linear} form~\cite{AHS:1985}. This is effectively
a sparse Boltzmann machine~\cite{AHS:1985} adopted to complex network
topologies without hidden units. (Note that hidden units usually play a
crucial role in ordinary Boltzmann machines~\cite{AHS:1985}). We introduce
a temporal evolution mechanism as a persistent sampling process for such a
machine based on the conditional probabilities obtained via the joint
probability, and generate a Markov chain of persistently sampled
state configurations to form the state transition time series for each node.
We call our model a \emph{sparse dynamical Boltzmann machine} (SDBM).

\begin{figure}
\centering
\includegraphics[width=\linewidth]{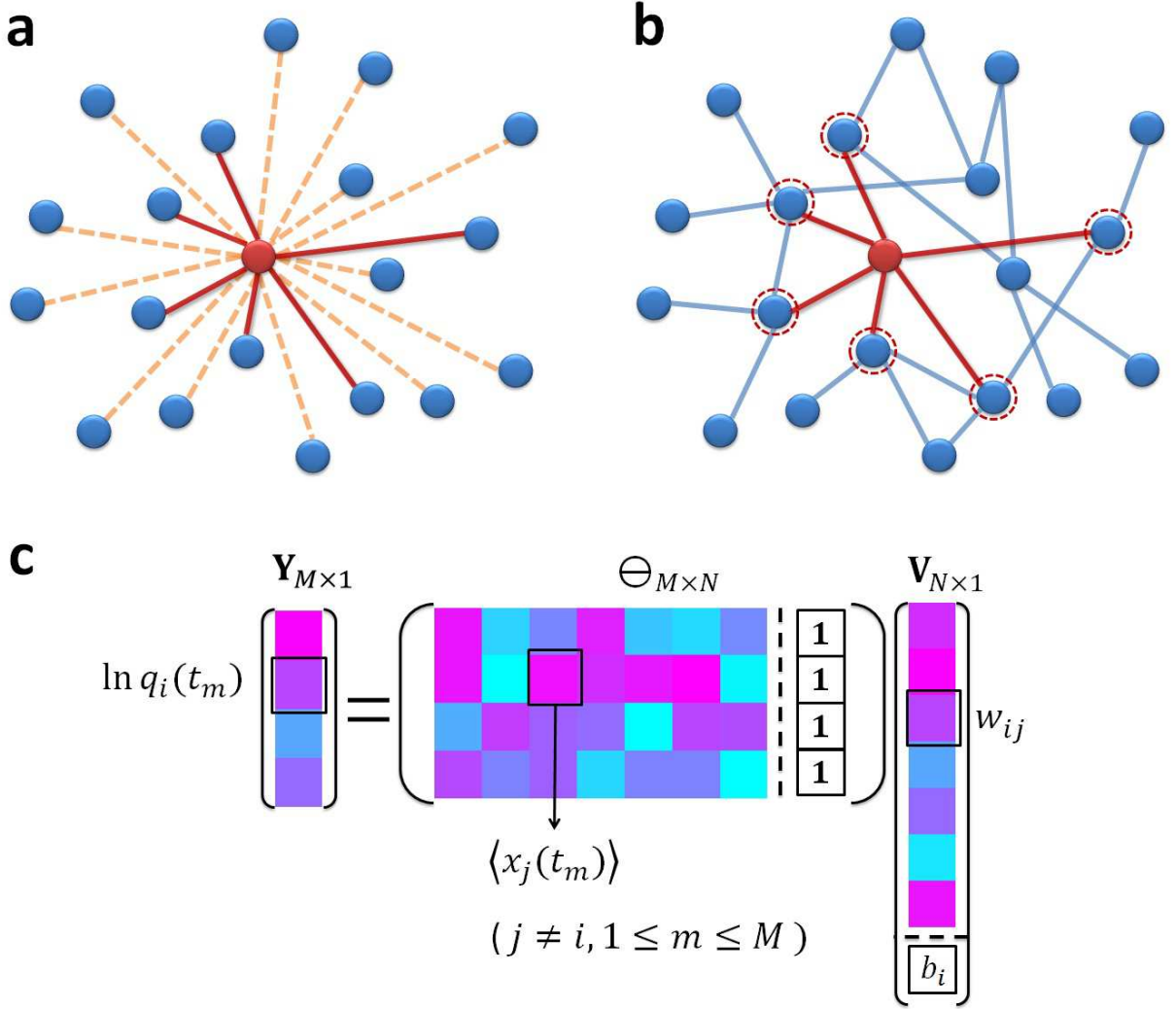}
\caption{ {\textbf{A schematic illustration of reconstruction of network 
structure using SDBM}. 
(a) Reconstruction of the local connection structure of the 
red node in a network of $20$ nodes. The connections of the network are
assumed to be unknown, so the red node can potentially be connected with 
any other node, as shown by the orange dashed links. Executing compressive 
sensing [(c)] and the K-means clustering [Fig.\ref{fig:kmeans}] for this 
node based on time series leads to its true connection structure, marked 
by the red links. The link weights and the nodal bias are represented by 
the vector $\mathbf{V}_{N\times1}$ in (c). (b) All true connections of the
network are recovered through the process in (a) for each node. The Markov 
blanket of the red node in (a) consists of all its nearest neighbors, 
indicated by the nodes with dashed red circles. (c) A schematic illustration 
of the compressive sensing framework for structural reconstruction as 
in Eq.~(\ref{eq:final_Eq}).}}
\label{fig:schematic1}
\end{figure}

For a dynamical process on complex networks, such as epidemic
spreading or evolutionary game dynamics, the state of each
node at the next time step is determined by the probability conditioned
on the current states of its neighbors (and its own state in some cases).
There is freedom to manipulate the conditional probabilities
that dictate the system behavior in the immediate future by adjusting the
values of its parameters, i.e., the weights and biases. A basic question
is then, for an SDBM, is it possible to properly choose these parameters
so that the conditional probabilities so produced are identical or nearly
identical to those of the network dynamical process with each given observed
system state configuration? If the answer is affirmative, the SDBM can
serve as a dynamics approximator of the original system, and the
approximated conditional probabilities possess predictive power for the
system state at the next time step. When such an SDBM is found for
many types of dynamical process on complex networks, it effectively serves
as a universal dynamics approximator. Moreover, if the detailed statistical
properties of the state configurations can be reproduced in the long time
limit, that is, if the time series generated by this SDBM are statistically
identical or nearly identical to those from the original system, then
the SDBM will be a generative model of the observed data (in the language of
machine learning), which is potentially capable of long term prediction.

When a dynamics approximator exists for the dynamics on a
complex network, the topology of the SDBM is nothing but that of the
original network, providing a simultaneous solution to the problem of
network structure reconstruction. Previous works on the inverse static
or kinetic Ising problems led to methods of reconstruction for Ising
dynamics by maximizing the data likelihood (or pseudo-likelihood) function
via the gradient descent approaches~\cite{T:1998,
CM:2011,AE:2012,NB:2012,R:2012,RH:2011,MJ:2011,SBD:2011,Z:2012,ZAAHR:2013,
BLHSB:2013,DR:2014}. Instead of adopting these approaches, as a part of
our methodology to extract the network structure, we articulate a
compressive sensing~\cite{CRT:2006a,CRT:2006b,Donoho:2006,Baraniuk:2007,
CW:2008,Romberg:2008} based approach, whose working power has been
demonstrated for a variety types of non-Ising type of dynamics on
complex networks~\cite{WYLKH:2011,WLGY:2011,SNWL:2012,SWL:2012,SLWD:2014,
SWFDL:2014,HSWD:2015}. By incorporating the K-means clustering algorithm 
into the sparse solution obtained from compressive
sensing, we demonstrate that nearly perfect reconstruction of the complex
network topology can be achieved. Using $14$ different types of dynamical
processes on complex networks, we find that, if the time series data
generated by these dynamical processes are assumed to be from
its equivalent SDBMs, the universal reconstruction framework is capable 
of recovering the underlying network structure for each type of original
dynamics with practically zero error. This represents solid and concrete
evidence that SDBM is capable of serving as a universal structural
estimator for complex networks. In addition to being able to precisely
reconstruct the network topology, the SDBM also allows the link weights
and nodal biases to be calculated with high accuracy.

Our method is fully automated and does not require any subjective parameter
choice.

\section{Results}

\subsection{SDBM as a network structural estimator}

For an SDBM of size $N$, the probability that the system is in a particular
binary state configuration
$\mathbf{X}_{N\times 1} = (x_1, x_2, \ldots, x_N)^{\mathrm{T}}$
is given by
\begin{equation} \label{eq:state_probability}
P(\mathbf{X}) = \frac{\exp(-E_{\mathbf{X}})}
{\sum_{\mathbf{X}}\exp(-E_{\mathbf{X}})},
\end{equation}
where $E_{\mathbf{X}}$ is the total energy of the network in $\mathbf{X}$:
\begin{equation} \label{eq:total_energy}
E_{\mathbf{X}} = \mathbf{X}^{\mathrm{T}}\cdot\mathbf{W}\cdot\mathbf{X}
= \sum_{i=1,i\neq j}^N\sum_{j=1,j\neq i}^N w_{ij} x_i x_j
- \sum_{i=1}^N b_i x_i,
\end{equation}
$x_i$ and $x_j$ are binary variables ($0$ or $1$) characterizing
the states of nodes $i$ and $j$, respectively, and $\mathbf{W}$ 
is a weighted matrix with its off diagonal
elements $w_{ij} = w_{ji}$ ($i,j = 1,\ldots, N, \ i\ne j$) specifying
the weight associated with the link between nodes $i$ and $j$. The
$i$th diagonal element of $\mathbf{W}$ is the bias
parameter $b_i$ for node $i$ ($i=1,\ldots,N$), which determines node
$i$'s preference to state $0$ or $1$. The total energy
$E_{\mathbf{X}}$ includes the interaction energies (the sum of all
$w_{ij}x_i x_j$ terms) and the nodes' self energies (the various
$b_i x_i$ terms). The partition function of the system is given by
\begin{equation} \label{eq:partition_function}
\mathcal{Z} = \sum_{\mathbf{X}}\exp(-E_{\mathbf{X}}),
\end{equation}
where the summation is over all possible $\mathbf{X}$ configurations.
The state of node $i$ at the next time step is determined by the states of
all the other nodes at the present time step, $\mathbf{X}^{\mathrm{R}}_i$,
through the following conditional probability
\begin{equation} \label{eq:CP}
P\{x_i(t+1)=1| \mathbf{X}^{\mathrm{R}}_i(t)\} =
\frac{P\{x_i(t+1)=1,\mathbf{X}^{\mathrm{R}}_i(t)\}}
{P\{x_i(t+1)=1,\mathbf{X}^{\mathrm{R}}_i(t)\} + P\{x_i(t+1)=0,
\mathbf{X}^{\mathrm{R}}_i(t)\}},
\end{equation}
where the two joint probabilities are given by
\begin{eqnarray}
\nonumber
& & P\{x_i(t+1) = 1,\mathbf{X}^{\mathrm{R}}_i(t)\} = \\ \nonumber
& & \ \ \ \ \ \ \frac{1}{\mathcal{Z}}
\exp{ [-\sum_{j=1,j\neq i}^N w_{ij}x_j(t) - b_i
 - \sum_{s=1, s\neq i}^N \sum_{j=1,j\neq i}^N w_{sj} x_s(t) x_j(t)-
\sum_{s=1, s \neq i}^N b_s x_s(t)]}, \\ \nonumber
& & P\{x_i(t+1)=0,\mathbf{X}^{\mathrm{R}}_i(t)\} = \frac{1}{\mathcal{Z}}
\exp{[-\sum_{s=1, s\neq i}^N \sum_{j=1, j\neq i}^N w_{sj} x_s(t) x_j(t)
- \sum_{s=1, s \neq i}^N b_s x_s(t)] }.
\end{eqnarray}
A Markov network defined in this fashion is in fact the kinetic Ising
model~\cite{RH:2011,MJ:2011,SBD:2011,Z:2012,ZAAHR:2013,BLHSB:2013,
DR:2014}. With the joint probabilities, the conditional
probability in Eq.~\eqref{eq:CP} becomes
\begin{equation} \label{eq:CP2}
P\{x_i(t+1)=1| \mathbf{X}^{\mathrm{R}}_i(t)\}
= \frac{1}{1 + \exp{[\sum_{j=1, j \neq i}^N w_{ij}x_j(t) + b_i]}}.
\end{equation}
We thus have
\begin{displaymath}
\ln{\left(\frac{1}{P\{x_i(t+1)=1| \mathbf{X}^{\mathrm{R}}_i(t)\}}-1 \right)}
= \sum_{j=1, j \neq i}^N w_{ij}x_j(t) +  b_i.
\end{displaymath}
Letting
$Q_i(t)\equiv\left(P\{x_i(t+1)=1| \mathbf{X}^{\mathrm{R}}_i(t)\}\right)^{-1}-1$,
we have
\begin{equation}
\ln{Q_i(t)} = \left( \begin{array}{ccccccccc}
x_1(t), & \ldots, & x_{i-1}(t), & x_{i+1}(t), & \ldots, & x_N(t), & 1\\
         \end{array}
\right)
\left( \begin{array}{cc}
w_{i1}\\
\vdots\\
w_{i(i-1)}\\
w_{i(i+1)}\\
\vdots\\
w_{iN}\\
b_i
      \end{array}
\right)
\end{equation}
For $M$ distinct time steps $t_1, t_2, \ldots, t_M$, we obtain the
following matrix form:
\begin{equation} \label{eq:matrix_form}
\left( \begin{array}{cc}
\ln{Q_i(t_1)} \\
\ln{Q_i(t_2)} \\
\vdots \\
\ln{Q_i(t_M)} \\
       \end{array}
\right)
= \left( \begin{array}{ccccccccc}
x_1(t_1), &\ldots, & x_{i-1}(t_1), &x_{i+1}(t_1), &\ldots, &x_N(t_1), & 1\\
x_1(t_2), &\ldots, & x_{i-1}(t_2), &x_{i+1}(t_2), &\ldots, &x_N(t_2), & 1\\
\vdots & \vdots & \vdots & \vdots & \vdots & \vdots & \vdots\\
x_1(t_M), &\ldots, & x_{i-1}(t_M), &x_{i+1}(t_M), &\ldots, &x_N(t_M), & 1\\
\end{array}
\right) \left( \begin{array}{cc}
w_{i1}\\
\vdots\\
w_{i(i-1)}\\
w_{i(i+1)}\\
\vdots\\
w_{iN}\\
b_i
               \end{array}
\right),
\end{equation}
which can be written concisely as
\begin{equation}
\mathbf{Y}_{M\times 1}=\mathcal{\Psi}_{M\times N}\cdot\mathbf{V}_{N\times 1},
\end{equation}
where the vector $\mathbf{Y}_{M\times1} \in R^{M}$ contains the values of
$\ln{Q_i(t)}$ for $M$ different time steps, the $M\times N$ matrix
$\mathcal{\Psi}_{M\times N}$ is determined by the states of all the nodes
except $i$, and the first $(N-1)$ components of the vector
$\mathbf{V}_{N\times1} \in R^N$ are the link weights between node $i$ and
all other nodes in the network, as illustrated in Fig.~\ref{fig:schematic1}(a), 
with its last entry being node $i$'s intrinsic bias.

Since, as shown in Fig.~\ref{fig:schematic1}(b), the conditional probability
$P\{x_i(t+1)=1| \mathbf{X}^{\mathrm{R}}_i(t)\}$ depends solely on the
state configuration of $i$'s nearest neighbors, or $i$'s
\emph{Markov blanket}~\cite{bishop:2006book,RN:2009book} at time $t$, 
identical configurations at other time steps imply identical conditional
probabilities. Thus, given time series data of the dynamical process, the
conditional probability can be estimated according to the law of large
numbers by averaging over the states of $i$ at all the time steps prior
to the neighboring state configurations becoming identical.
Note, however, that this probability needs to be conditioned on the state
configuration of the entire system except node $i$, i.e., on
$\mathbf{X}^{\mathrm{R}}_i(t)$, and the average of $x_i$ is calculated
over all the time steps $t_m+1$ satisfying
$\mathbf{X}^{\mathrm{R}}_i(t) = \mathbf{X}^{\mathrm{R}}_i(t_m)$. This
means that there can be a dramatic increase in the configuration size,
i.e., from $k_i$ (the degree of node $i$) to $N-1$ (the size of
the vector $\mathbf{X}^{\mathrm{R}}_i$), which can make the number of exactly
identical configurations too small to give any meaningful statistics. To
overcome this difficulty, we allow a small amount of dissimilarity
between $\mathbf{X}^{\mathrm{R}}_i(t) $ and $\mathbf{X}^{\mathrm{R}}_i(t_m)$
by introducing a tolerance parameter, $\Gamma$, to confine the
corresponding Hamming distances normalized by $N$. In particular, we assume 
$\mathbf{X}^{\mathrm{R}}_i(t) \approx \mathbf{X}^{\mathrm{R}}_i(t_m)$ 
if the relative difference between them is not larger than $\Gamma/N$. 
This averaging process leads to
{\small
\begin{eqnarray} \label{eq:final_Eq}
\left(\begin{array}{cc}
\ln{q_i(t_1)} \\
\ln{q_i(t_2)} \\
\vdots\\
\ln{q_i(t_M)} \\
      \end{array} \right)
= &\left( \begin{array}{lllllllll}
\langle x_1({t}_1)\rangle,&\ldots,&\langle x_{i-1}({t}_1)\rangle,&\langle x_{i+1}({t}_1)\rangle,&\ldots,&\langle x_N({t}_1)\rangle,&1\\
\langle x_1({t}_2)\rangle,&\ldots,&\langle x_{i-1}({t}_2)\rangle,&\langle x_{i+1}({t}_2)\rangle,&\ldots,&\langle x_N({t}_2)\rangle,&1\\
\vdots &\vdots&\vdots&\vdots&\vdots &\vdots&\vdots\\
\langle x_i({t}_M)\rangle,&\ldots,&\langle x_{i-1}({t}_M)\rangle,&\langle x_{i+1}({t}_M)\rangle,&\ldots,&\langle x_N({t}_M)\rangle,&1\\
          \end{array}\right)
\left( \begin{array}{cc}
w_{i1}\\
\vdots\\
w_{i(i-1)}\\
w_{i(i+1)}\\
\vdots\\
w_{iN}\\
b_i
       \end{array} \right)
\end{eqnarray}} 
where $q_i(t) \equiv \langle x_i({t}_1 + 1)\rangle^{-1} - 1$, with 
$\langle \cdot \rangle$ standing for the averaging over all instants of
time at which the condition
$\mathbf{X}^{\mathrm{R}}_i(t) = \mathbf{X}^{\mathrm{R}}_i(t_m)$ is met.
A schematic illustration of the whole process is presented in 
Fig.~\ref{fig:schematic1}, with Eq.~\eqref{eq:final_Eq} shown in panel (c).

For a complex network, the degree of a typical node is small compared
with the network size. For node $i$, the link weights are nonzero only
for the connections with the immediate neighbors. The vector
$\mathbf{V}_{N\times1}$ is thus typically sparse with the majority
of its elements being zero. The sparsity property renders applicable
compressive sensing~\cite{CRT:2006a,CRT:2006b,Donoho:2006,Baraniuk:2007,
CW:2008,Romberg:2008}, through which an $N$ dimensional sparse vector
can be reconstructed via a set of $M$ measurements, for $M \ll N$.
By minimizing the $L_1$ norm of $\mathbf{V}_{N\times 1}$, i.e.,
$\| \mathbf{V}_{N\times 1} \|_1 = \sum_{j=1, j\neq i}^N |w_{ij}| + |b_i|$,
subject to the constraint $\mathbf{Y}_{M\times 1} = 
\mathcal{\Psi}_{M\times N} \cdot \mathbf{V}_{N\times 1}$, 
we can reconstruct $\mathbf{V}_{N\times 1}$ to
obtain the connection weights between node $i$ and all other nodes in
the network. One tempts to hope that, applying the procedure to each node
would lead to the complete weighted adjacency matrix, $\mathbf{W}$.

To gain insights, we test the structural estimation method for an
SDBM itself [Figs.~\ref{fig:solution}(a-c)] and three different
types of dynamical processes [Figs.~\ref{fig:solution}(d-f)] by feeding
the time series data generated by the SDBM into the framework and comparing
the reconstructed SDBM with the original machine. The time instants
$t_1, t_2, \ldots, t_M$ needed to construct Eq.~\eqref{eq:final_Eq} are
chosen randomly from $T$ time instants in total. For each node, 
compressive sensing is implemented a multiple of times to
obtain the averaged relevant quantities. As shown in Fig.~\ref{fig:solution},
the averaged solutions from the compressive sensing algorithm
appear in sharp peaks at places that correspond to the existent links,
despite the large differences among the values. This means that, while
most existent links can be predicted against the null links, the
accuracy of the solution so obtained is not sufficient for the actual
element values of $\mathbf{V}_{N\times 1}$ to be determined. For the
null links, the corresponding solutions generally appear as a noisy
background. For the ideal case where $w_{ij}$ is zero, the background
noise can be quite large especially for the large degree nodes, as shown
in Figs.~\ref{fig:solution}(b) and \ref{fig:solution}(d). Further
calculations indicate that averaging a larger number of simulation
runs can suppress the background noise to certain extent, but it cannot
be eliminated and may become quite significant for various types of
dynamics (Sec.~\ref{subsec:SDBM_universality}).

In previous works on reconstruction of complex networks based on
stochastic dynamical correlations~\cite{WCHLH:2009,RWLL:2010} or
compressive sensing~\cite{WLGY:2011,SWFDL:2014}, the existent (real)
links can be distinguished from the nonexistent links by setting
a single threshold value of certain quantitative measure. The
success relies on the fact that the dynamics at various nodes are
of the same type, and the reconstruction algorithm is tailored
toward the specific type of dynamical process. Our task is significantly
more challenging as the goal is to develop a universal system (or
machine) to replicate a diverse array of dynamical processes based
on data. As can be seen from Figs.~\ref{fig:solution}(a-c), for
compressive sensing based reconstruction, the computational criteria
to distinguish existent from nonexistent links differ substantially
for different dynamics in terms of quantities such as the solution
magnitude, peak value distribution, and background noise intensity.
As a result, a more elaborate and sophisticated procedure is
required for determining the threshold for each particular case,
suggesting that a straightforward application of compressive sensing
cannot lead to a universal reconstruction algorithm. One may also
regard the solutions of the existent links as a kind of \emph{extreme
events}~\cite{KSA:2011,CHL:2014,CHZESL:2015} superimposed on top of the 
random background, but it is difficult to devise a universal criterion 
to determine if a peak in the distribution of the quantitative measure 
represents the correct extreme event corresponding to an existent link.

Through extensive testing, we find that a previously developed
unsupervised clustering measure, the K-means~\cite{bishop:2006book,
RN:2009book}, possesses the desired traits that can be exploited,
in combination with compressive sensing, to develop a universal 
reconstruction machine (see {\bf Methods} for
details). As we will show, K-means can serve as a base for a
highly effective structural estimator for various types of dynamics
on networks of distinct topologies. Depending on the specific
combination of the network topology and dynamics, the reconstruction
accuracies vary to certain extent but are acceptable.
Since the compressive sensing operation is node specific, the
solutions obtained separately from different nodes may give
conflicting information as to whether there is an actual link
between the two nodes, requiring a proper resolution scheme.
We develop such a scheme based on node degree consistency
(see {\bf Methods}). Our universal reconstruction machine thus
contains three main components: compressive sensing, K-means,
and conflict resolution. We shall demonstrate that the machine
can separate the true positive solutions from the noisy background
with high success rate, for all combinations of the nodal dynamics and
the network topology tested.

\subsection{Universality of SDBM as a network structural estimator}
\label{subsec:SDBM_universality}

\begin{figure}
\centering
\includegraphics[width=\linewidth]{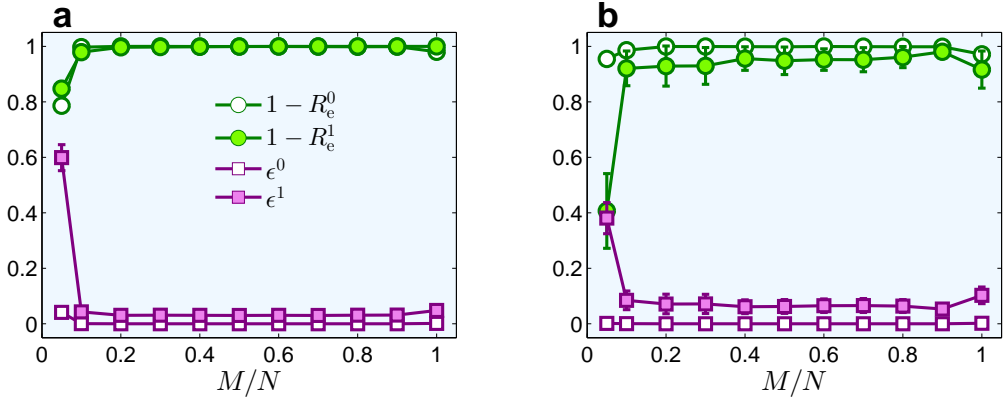}
\caption{\textbf{Structural estimation success rates and parameter
estimation errors versus the normalized data amount.}
For SDBM dynamics on (a) ER-random and (b) BA scale-free networks,
the success rates of detection of existent links and identification
of nonexistent links, $1-R^1_\mathrm{e}$ (green circles) and
$1-R^0_\mathrm{e}$ (white circles), respectively, versus the
normalized data amount $M/N$ with error bars. The absolute values
of the errors in estimating the link weights and the nodal biases
for the existent links (purple squares) and non-existent
links (white squares) are also shown. Most data points have quite
small error bars, indicating the performance stability
of our method. High performance structural and parameter estimations
are achieved insofar as $M/N$ exceeds about $10\%$.}
\label{fig:measurements}
\end{figure}

Figure~\ref{fig:measurements} demonstrates the performance of network
structural estimation for SDBM dynamics on random and scale-free
networks in terms of the data amount,
characterized by the number of measurements $M$ normalized by $N$,
for SDBMs built up for systems with different original topologies.
We define $R_{\mathrm{e}}^1$ and $R_{\mathrm{e}}^0$ as the estimation
error rates for the existent and non-existent links, namely, the false
positive and the false negative rates, respectively. We see that,
for a wide range of the values of $M/N$, the success rates of the
existent and non-existent links, $1-R_{\mathrm{e}}^1$ and
$1-R_{\mathrm{e}}^0$, respectively, are nearly $100\%$ for homogeneous
network topology. For heterogeneous (scale-free) networks, the success
rate $1-R_{\mathrm{e}}^1$ tends to be slightly lower than $100\%$, due to
the violation of the sparsity condition for hub nodes, leading to
the difficulty to distinguish the peaks in the distribution of
the compressive sensing solutions from the noisy background. For $M/N=1$,
the success rates are reduced slightly, due to the non-zero dissimilarity
tolerance $\Gamma$ that makes the values of $\mathbf{X}_i^\mathrm{R}$
at different time steps indistinguishable and introduces linear
dependence into Eq.~\eqref{eq:final_Eq}. As a result, the accuracy of
compressive sensing solutions is compromised to certain extent. 
Generally, high reconstruction precision is guaranteed for almost any 
choice of $M$ insofar as it is not too close to $0$ and $N$, and this 
feature makes our SDBM framework free of any subjective parameters.

\begin{figure}
\centering
\includegraphics[width=\linewidth]{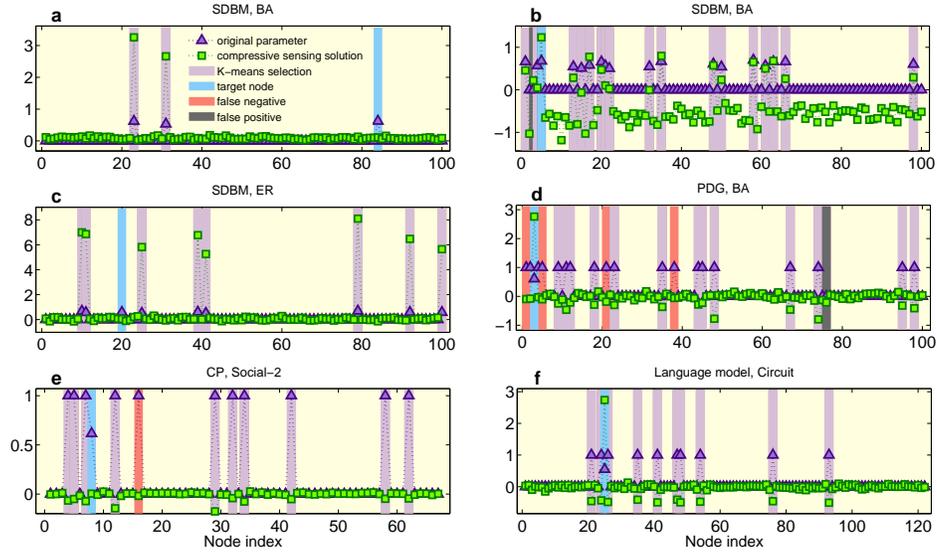}
\caption{\textbf{Solution examples of compressive sensing and 
K-means clustering results.}
The compressive sensing solutions (green squares) are compared,
element-wise, with the true values of the original connection
vectors (dark purple triangles), where the x-axis is the index of nodes
with different degree values for different network topologies and dynamics
types. The target nodes and the correctly detected existent connections
via the K-means are marked as light blue and light purple bars, respectively.
The false negative and the false positive estimations are marked as red
and gray bars, respectively. For all cases, the measurement amount 
is $M=0.4 N$. The restriction on Hamming distances is set to
be $\Gamma=0.35$ for all examples, and a small change in $\Gamma$ would
not affect the qualitative results. (a) A node with the smallest degree
$k=2$ in a BA scale-free network of size $N=100$ and average degree
$\langle k \rangle = 4$ with SDBM dynamics. (b) The node of the largest
degree $k=18$ in the same network. (c) A node of degree $k=8$ in an ER
random network of size $N=100$ and average degree $\langle k \rangle = 4$
with SDBM dynamics. (d) The node with the largest degree $k=19$ in a
BA scale-free network of size $N=100$ and average degree
$\langle k \rangle = 4$ with prisoner's dilemma game (PDG) dynamics.
(e) A node of degree $k=11$ in a real world social network of size
$N=67$ subject to CP dynamics. (f) A node of degree $k=10$ in a
real world electrical circuit network of size $N=122$ with Language
model dynamics. In (d-f), the element values of the original connection
vector are set to unity since no true weight values are given. For
cases (a,b,c,f), there are no false negative detections, i.e., all
existent links have been successfully detected. For cases (a,c,e),
there are no false positive detections, i.e., no non-existent links
are mistaken as existent ones. A detailed description of the dynamical
processes is given in Tab.~\ref{tab:topo1}.}
\label{fig:solution}
\end{figure}

\begin{table}
\begin{center}
\caption{\textbf{Description of the 14 dynamical processes on model
networks and the conditional probabilities.} The quantity
$P_i^{0\rightarrow 1}$ or $P_i^{1\rightarrow 0}$ is the probability
that the state of node $i$ (of degree $k_i$) becomes $1$ at the next
time step when its current state is $0$, or vice versa, conditioned
on its current neighboring state configuration. The number of $i$'s neighbors
in state $1$ is $n_i$. (See Supplementary Note 1 for a detailed
description of the dynamics and parameters for estimating the conditional
probabilities $P_i^{0\rightarrow1}$ and $P_i^{1\rightarrow0}$.)}
\label{tab:topo1}
\setlength{\extrarowheight}{3pt}
\noindent\makebox[\textwidth]
{%
\centering
\begin{tabularx}{\linewidth}{c c c c}
\cline{1-4}
Category & Dynamics Type & $P_i^{0\rightarrow1}$ & $P_i^{1\rightarrow0}$ \\
\cline{1-4} \\
\multirow{1}{*}{\begin{minipage}{0.1in} I \end{minipage}} & SDBM & $\frac{1}{1 + \exp[\sum_{j=1, j \neq i}^N w_{ij}x_j(t) + b_i]}$ & $\frac{\exp[\sum_{j=1, j \neq i}^N w_{ij}x_j(t) + b_i]}{1 + \exp[\sum_{j=1, j \neq i}^N w_{ij}x_j(t) + b_i]}$ \\
\\
\multirow{1}{*}{\begin{minipage}{0.2in} II \end{minipage}} & Ising Glauber~\cite{G:1963}  & $\frac{1}{1+\exp[\frac{2J}{\kappa}(k_i - 2 n_i)]}$  &  $\frac{\exp[\frac{2J}{\kappa}(k_i - 2 n_i)]}{1+\exp[\frac{2J}{\kappa}(k_i - 2 n_i)]}$ \\
& SQ-SG~\cite{GWW:2007}  &  $\frac{1}{1+\exp[(rk_i-n_i)/\kappa]}$ & $\frac{\exp[(rk_i-n_i)/\kappa]}{1+\exp[(rk_i-n_i)/\kappa]}$ \\
& SQ-PDG~\cite{GWW:2007} & $\frac{1}{1+\exp[(b-1)(k_i-n_i)/\kappa]}$ & $\frac{\exp[(b-1)(k_i-n_i)/\kappa]}{1+\exp[(b-1)(k_i-n_i)/\kappa]}$ \\
  \\
  \multirow{1}{*}{\begin{minipage}{0.2in} III \end{minipage}} & Minority game~\cite{CZ:1997,CMZ:2013,PBC:2000,ZWZYL:2005,HZDHL:2012}  & $\frac{k_i - n_i}{k_i}$ & $\frac{n_i}{k_i}$ \\
&  Voter~\cite{L:book,O:1992} & $\frac{n_i}{k_i}$ & $\frac{k_i - n_i}{k_i}$ \\
&\multirow{4}{*}{\begin{minipage}{1.2in}\centering Majority vote~\cite{L:book,O:1992}\end{minipage}}
  &   \multirow{1}{*}{\begin{minipage}{1.2in} \begin{eqnarray}   \left\{
  \begin{aligned} &Q & \text{if   } n_i < k_i/2\\ \nonumber
  &1/2 & \text{if   } n_i = k_i/2 \\ \nonumber  &1-Q & \text{if   } n_i > k_i/2 \nonumber \end{aligned}
  \right.\end{eqnarray}
  \end{minipage}} & \multirow{3}{*}{\begin{minipage}{1.2in} \begin{eqnarray}   \left\{
  \begin{aligned} &1-Q & \text{if   } n_i < k_i/2\\ \nonumber
  &1/2 & \text{if   } n_i = k_i/2 \\ \nonumber  &Q & \text{if   } n_i > k_i/2 \nonumber \end{aligned}
  \right.\end{eqnarray}
  \end{minipage}} \\ \\ \\
\\
\multirow{1}{*}{\begin{minipage}{0.2in} IV \end{minipage}} & Link-update voter~\cite{BBV:book,CFL:2009}  & $\frac{n_i}{\langle k \rangle}$ & $\frac{k_i - n_i}{\langle k \rangle}$ \\
  & Language model\cite{VCS:2010,AS:2003}  &  $S(\frac{n_i}{k_i})^\alpha$ & $(1-S) (\frac{k_i-n_i}{k_i})^\alpha$ \\
  & Kirman~\cite{K:1993,ALW:2005,ALW:2005}  & $c_1 + d n_i$  & $c_2 + d (k_i - n_i)$ \\
  \\
  \multirow{1}{*}{\begin{minipage}{0.2in} V \end{minipage}} & CP~\cite{CP:2006,VM:2009}  & $\frac{n_i}{k_i} \lambda_i$ & $\mu_i$ \\
  & SIS~\cite{B:book,AMay:book,PV:2001a,PV:2001b}  & $1-(1-\lambda_i)^{n_i}$ & $\mu_i$ \\
  \\
  \multirow{1}{*}{\begin{minipage}{0.2in} VI \end{minipage}}
  & SG~\cite{S:1986}  & * & * \\
  & PDG~\cite{AH:1981,ST:1998} & * & * \\
\cline{1-4}
\end{tabularx} }
\end{center}
\end{table}

\begin{table}
\begin{center}
\caption{\textbf{Reconstruction error rates.}
For the same dynamical processes in Tab.~\ref{tab:topo1}, the error
rates (in percentages) in uncovering the existent and non-existent
links, $R_\mathrm{e}^1$ and $R_\mathrm{e}^0$, respectively. ER-random
and BA scale-free networks of size $N=100$ and average degree 
$\langle k\rangle=4$ are used as the underlining network supporting 
the various dynamical processes. The normalized number of measurements 
used in compressive sensing is $M/N=0.4$ for all cases.}
\label{tab:error_rates}
\setlength{\extrarowheight}{3pt}
\noindent\makebox[\textwidth]
{%
\centering
\begin{tabularx}{\linewidth}{c c c c}
\cline{1-4}
Category & Dynamics Type & $R_\mathrm{e}^0$ / $R_\mathrm{e}^1$ (\%, ER) & $R_\mathrm{e}^0$ / $R_\mathrm{e}^1$ (\%, BA) \\
\cline{1-4}
\multirow{1}{*}{\begin{minipage}{0.1in} I \end{minipage}} & SDBM & 0.0 / 0.1 & 0.1 / 3.4 \\
\multirow{1}{*}{\begin{minipage}{0.2in} II \end{minipage}} & Ising Glauber~\cite{G:1963} & 0.0 / 0.0 & 0.0 / 0.0 \\
& SQ-SG~\cite{GWW:2007}  & 0.0 / 0.0 & 0.0 / 3.2 \\
& SQ-PDG~\cite{GWW:2007} & 0.0 / 0.0 & 0.3 / 3.3 \\
\multirow{1}{*}{\begin{minipage}{0.2in} III \end{minipage}} & Minority game~\cite{CZ:1997,CMZ:2013,PBC:2000,ZWZYL:2005,HZDHL:2012} & 0.0 / 0.6 & 0.0 / 0.2 \\
& Voter~\cite{L:book,O:1992} & 0.0 / 0.0 & 0.0 / 0.0 \\
& Majority vote~\cite{L:book,O:1992} & 0.0 / 0.0 & 0.0 / 0.1 \\
\multirow{1}{*}{\begin{minipage}{0.2in} IV \end{minipage}} & Link-update voter~\cite{BBV:book,CFL:2009} & 0.0 / 0.0 & 1.8 / 3.6  \\
& Language model\cite{VCS:2010,AS:2003}  & 0.0 / 0.0 & 0.0 / 0.4 \\
& Kirman~\cite{K:1993,ALW:2005,ALW:2005} & 0.0 / 2.3 & 0.0 / 3.8  \\
\multirow{1}{*}{\begin{minipage}{0.2in} V \end{minipage}} & CP~\cite{CP:2006,VM:2009} & 0.0 / 1.4 & 0.0 / 0.4  \\
& SIS~\cite{B:book,AMay:book,PV:2001a,PV:2001b} & 0.3 / 5.2 & 0.1 / 15.6  \\
\multirow{1}{*}{\begin{minipage}{0.2in} VI \end{minipage}}
& SG~\cite{S:1986} & 0.0 / 1.7 & 0.0 / 9.4  \\
& PDG~\cite{AH:1981,ST:1998} & 0.1 / 9.6 & 0.1 / 9.8 \\
\cline{1-4}
\end{tabularx} }
\end{center}
\end{table}

Our working hypothesis is that, for a networked system with a certain type
of nodal dynamics, there exists an equivalent SDBM. Reconstructing the
structure of the SDBM would simultaneously give the topology of the
original networked dynamical system. Accordingly, the time series data
generated from the original system can be used to reveal its underlining
interaction structures through the corresponding SDBM. In particular,
we directly feed the original time series into the framework of compressive
sensing and K-means for SDBM to generate the network structure, and test
whether the structure generated by the SDBM based reconstruction represents
that of the original network. The similarity can be quantified by the
error rates of the existent and non-existent links. Table~\ref{tab:topo1}
list the 14 dynamical processes on model networks and the underlying
conditional probabilities, and Tab.~\ref{tab:error_rates} shows the
performance in terms of the percentage error rates $R_\mathrm{e}^0$ and
$R_\mathrm{e}^1$. The $14$ types of dynamical processes are taken from
the fields of evolutionary game theory, opinion dynamics, and spreading
processes, covering a number of focused research topics in complex
networks. Strikingly, for all the dynamics with diverse properties,
we find that that, for each and every dynamics-network combination,
zero or nearly zero error rates are obtained for both the existent and
non-existent links, revealing a strong similarity between the original
networks and the ones generated from SDBM, regardless of the type of
dynamics. The nonzero error rates in Tab.~\ref{tab:error_rates} come
mainly from the high degree nodes. Consequently, as indicated in
Tab.~\ref{tab:error_rates}, the reconstruction accuracy for networks of
homogeneous topology is generally higher than that for heterogeneous networks.

For certain types of evolutionary game dynamics, especially for the
snowdrift game (SG)~\cite{S:1986} and the prisoner's dilemma game
(PDG)~\cite{AH:1981,ST:1998} with the Fermi updating
rule~\cite{ST:1998}, information about the state configuration of the
second nearest neighbors is required to calculate the payoffs of the
first nearest neighbors. In such a system, the next move of a target
node is determined by comparing its payoff with those of its neighbors.
This implies that, using solely the state configuration information of
the Markov blanket, without the aid of the payoff information that
requires the state information of the second nearest neighbors, is
insufficient to determine the state of the target node into the immediate
future, rendering inapplicable our SDBM based reconstruction.
Contrary to this intuition, we find that for both SG and PDG,
high reconstruction accuracy can be achieved, as shown in
Tab.~\ref{tab:error_rates}. Recall that our reconstruction method is
formulated based on the independence assumption of Markov networks,
i.e., in order to reconstruct the local structure of a target node,
it should be completely independent of the rest of the system when
the configuration of its Markov blanket is given. The results in
Tab.~\ref{tab:error_rates} indicate that our SDBM based algorithm
performs better than expected in terms of the reconstruction accuracy.
In fact, the independence assumption can be made to hold by adopting
a self-questioning (SQ) based updating rule (see Supplementary Note 1
for details). In this case, excellent reconstruction accuracy is
obtained, as can be seen from Tab.~\ref{tab:error_rates}.

The structural estimation results reported so far are based on model
network topologies. Real world complex networks have also been used to
test our framework, with results shown in Figs.~\ref{fig:solution}(e,f)
and Supplementary Notes 2 and 3.
For various combinations of the network topology and dynamical process,
high reconstruction accuracy is achieved, where for a number of cases
the error rates are essentially zero. There are a few special cases
where the errors are relatively large, corresponding to situations where
the globally frozen or oscillating states dominate the dynamical process
so that too few linearly independent measurements can be obtained.
Overall, the equivalent SDBM correspondence holds and 
our reconstruction scheme for real world networks is effective.

\subsection{Parameter estimation scheme}

\begin{figure}
\centering
\includegraphics[width=\linewidth]{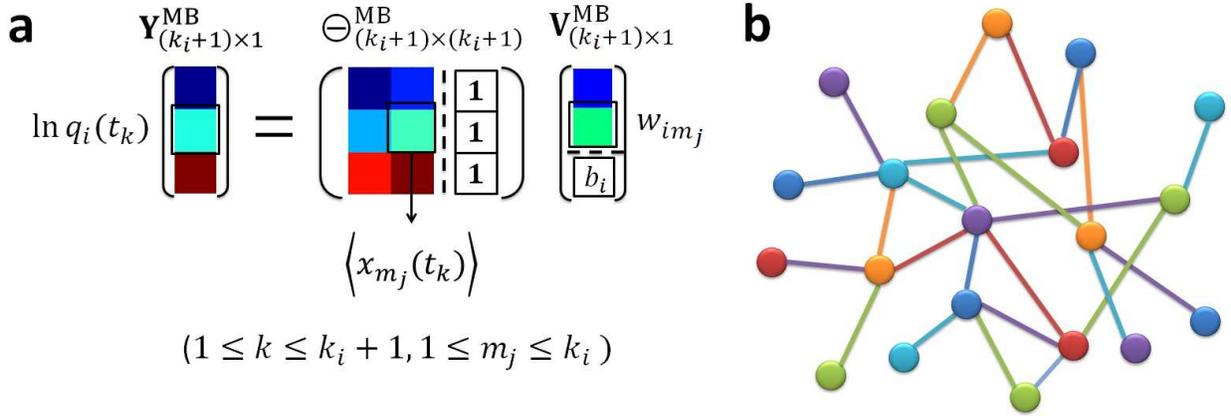}
\caption{ \textbf{A schematic illustration of SDBM parameter estimation.}
(a) For the SDBM of the same system in Fig.~\ref{fig:schematic1}(a,b), the 
corresponding parameter estimation framework as in Eq.~(\ref{eq:final_Eq2}). 
The connections of the network have already been reconstructed, so the 
entries in Eq.~(\ref{eq:final_Eq2}) can be obtained from a node's Markov 
blanket [Fig.~\ref{fig:schematic1}(b)]. The calculation is implemented for 
each node in the system. (b) The values of the link weights and the nodal 
biases (different colors) for the corresponding SDBM are calculated via 
the parameter estimation scheme in (a) and the degree guided solution 
substitution operation (see {\bf Methods}).}
\label{fig:schematic2}
\end{figure}

From the reconstructed network structure, any node $i$'s $k_i$ immediate
neighbors, $m_{1}$, $m_{2}$, $\ldots$, $m_{k_i}$, and their state configuration,
$\mathbf{X}^{\mathrm{MB}}_i = (x_{m_{1}}, x_{m_{2}}, \ldots,
x_{m_{k_i}})^\mathrm{T}$, can be identified. Since the probability that
$i$'s state is $1$ at the next time step depends only on $i$'s immediate
neighbors, $\mathbf{X}^{\mathrm{R}}_i(t)$ in Eq.~\eqref{eq:CP2} can be
simplified to $\mathbf{X}^{\mathrm{MB}}_i(t)$, and Eq.~\eqref{eq:CP2} can
be written as
\begin{equation}
P\{x_i(t+1)=1|\mathbf{X}^{\mathrm{R}}_i(t)\}
= P\{x_i(t+1)=1|\mathbf{X}^{\mathrm{MB}}_i(t)\}
= \frac{1}{1+\exp[\sum_{m=1}^{k_i} w_{im} x_m(t) + b_i]}.
\end{equation}
Accordingly, Eq.~\eqref{eq:final_Eq} can be simplified as
\begin{displaymath}
\mathbf{Y}^{\mathrm{MB}}_{(k_i+1) \times 1}
= \mathcal{\Psi}^{\mathrm{MB}}_{(k_i+1) \times (k_i+1)} \cdot
\mathbf{V}^{\mathrm{MB}}_{(k_i+1) \times 1},
\end{displaymath}
i.e.,
\begin{eqnarray} \label{eq:final_Eq2}
\left(
\begin{array}{cc}
\ln{q_i(t_1)} \\
\ln{q_i(t_2)} \\
\vdots\\
\ln{q_i(t_{k_i+1})} \\
\end{array}
\right)
= &\left(
\begin{array}{ccccccccc}
\langle x_{m_1}({t}_1)\rangle,  & \langle x_{m_2}({t}_1)\rangle, \ldots, & \langle x_{m_{k_i}}({t}_1)\rangle, & 1\\
\langle x_{m_1}({t}_2)\rangle,  & \langle x_{m_2}({t}_2)\rangle, \ldots, & \langle x_{m_{k_i}}({t}_2)\rangle, & 1\\
\vdots & \vdots & \vdots & \vdots\\
\langle x_{m_1}({t}_{k_i+1})\rangle,  & \langle x_{m_1}({t}_{k_i+1})\rangle, \ldots, & \langle x_{m_{k_i}}({t}_{k_i+1})\rangle, & 1\\
\end{array}
\right)
\left(
\begin{array}{cc}
w_{im_1}\\
w_{im_2}\\
\vdots\\
w_{im_{k_i}}\\
b_i
\end{array}
\right),
\end{eqnarray}
where the $k_i+1$ linear equations uniquely solve
$w_{im_1}, \ldots, w_{im_{k_i}}$ and $b_i$ simply via
\begin{displaymath}
\mathbf{V}^{\mathrm{MB}}_{(k_i+1) \times 1}
= [\mathcal{\Psi}^{\mathrm{MB}}_{(k_i+1) \times (k_i+1)}]^{-1} \cdot
\mathbf{Y}^{\mathrm{MB}}_{(k_i+1) \times 1}.
\end{displaymath}
Our parameter estimation formulation is illustrated schematically in 
Fig. \ref{fig:schematic2}(a).

For a particular neighboring state configuration of node $i$,
$\mathbf{X}^{\mathrm{MB}}_i(t)$, its occurring frequency determines
the precision in the estimation of $P\{x_i(t+1)|\mathbf{X}^{\mathrm{MB}}_i(t)\}
\simeq \langle x_i(t + 1)\rangle$, which in turn determines the solution
precision of Eq.~\eqref{eq:final_Eq2}. The occurrence of different
neighboring state configurations for the same node may differ dramatically.
Furthermore, the accuracy in estimating the probability
$P\{x_i(t+1)|\mathbf{X}^{\mathrm{MB}}_i(t)\}$ depends on the node degree
due to the increasing difficulty in finding exactly the same configurations
for larger degree nodes. Overcoming the estimation difficulty is a highly
non-trivial problem. Exploiting the available data further, we develop
a degree guided solution-substitution operation to cope with this difficulty,
where sufficient estimation precision is guaranteed for most cases (see
{\bf Methods}). A SDBM with estimated parameters is 
schematically shown in Fig.~\ref{fig:schematic2}(b).

\subsection{Universal dynamics approximator}

In Table~\ref{tab:topo1}, the complex network dynamics are categorized
in terms of the specific probabilities $P_i^{0\rightarrow1}$
and $P_i^{1\rightarrow0}$. In particular, category I is for the SDBM dynamics
and category II contains dynamical processes with $P_i^{0\rightarrow1}$ and
$P_i^{1\rightarrow0}$ having a mathematical form similar to that of
the SDBM. For category III, the forms of $P_i^{0\rightarrow1}$ and
$P_i^{1\rightarrow0}$ for the three types of dynamics are quite different.
Despite the differences among the dynamical processes in the three
categories, they share a key property: i.e.,
$P_i^{0\rightarrow1} + P_i^{1\rightarrow0}=1$, which plays a critical
role in implementing the parameter estimation algorithm. For the 
processes in categories IV and V, we have
$P_i^{0\rightarrow1} + P_i^{1\rightarrow0} \neq 1$, where
$P_i^{1\rightarrow0}$ is a time-invariant constant for the nodal dynamics
in category V.

We first test our parameter estimation scheme using the state time
series generated by the SDBM, which can be validated
through a direct comparison of the estimated parameter values
with their true values, as shown in Figs.~\ref{fig:solution}(a-c). However,
since the parameters only affect the system collectively (not individually)
in the form of the product summation [Eq.~\eqref{eq:CP2}] and our goal is
to assess the predictive power, we introduce an alternative validation
scheme. For each particular system configuration, the acceptable parameter
estimation results would serve as a base to generate another SDBM with
identical conditional probabilities for each node at each time step, as
compared with those of the original system. (A visual comparison between 
the conditional probabilities of the original system and that generated 
via the reconstructed SDBM is presented in Supplementary Note 4).
Figure~\ref{fig:all_error_distr}(a) shows the error
distributions of the estimated conditional probability time series, where
an overwhelmingly sharp peak occurs at $0$, indicating an excellent
agreement between the estimated and the true parameter values.

For typical dynamical processes on complex networks, our goal is to
find the equivalent SDBMs whose true parameter values are not available.
In this case, the performance of the parameter estimation scheme can be
assessed through the reconstructed conditional probabilities. Based on the
recovered network structure, the time series of the network dynamics are
fed into the parameter estimation scheme, and the link weights and the
nodal biases are obtained to form a system obeying the SDBM dynamics.
Corresponding to the categories in Table~\ref{tab:topo1}, the distributions
of the estimation errors between the original conditional probability
$P\{x_i(t+1)=1| \mathbf{X}^{\mathrm{R}}_i(t)\}$ and the one generated by
the newly constructed SDBM are shown in Fig.~\ref{fig:all_error_distr},
where panels (a-f) show the error distributions corresponding to the
six types of dynamics in categories II and III in Table~\ref{tab:topo1},
respectively. In each case, a sharp peak at zero dominates
the distribution, indicating the equivalence of the reconstructed
SDBM to the original dynamics. Given a particular type of complex network
dynamics, the SDBM resulted from our structural and parameter estimation
framework is indeed equivalent to the original dynamical system. The limited
amount of data obtained from the original system renders important state
prediction of the system, a task that can be accomplished by taking
advantage of the equivalence of the SDBM to the original system in the
sense that the SDBM produces approximately equal state transition
probabilities in the immediate future, given the current system
configuration. The SDBM thus possesses a significant predictive power
for the original system. Regardless of the type of the dynamical
process, insofar it satisfies the relation $P_i^{0\rightarrow1}
+ P_i^{1\rightarrow0} =1$, the reconstructed SDBM can serve as a universal
dynamics approximator.

For an SDBM, the relation $P_i^{1\rightarrow1} = 1- P_i^{1\rightarrow0}
= P_i^{0\rightarrow1}$ holds in general. However, for the dynamical
processes in categories IV and V, we have
$P_i^{1\rightarrow1} \neq 1- P_i^{1\rightarrow0} = P_i^{0\rightarrow1}$
so that a single SDBM is not sufficient to fully characterize the dynamical
evolution. Our solution is to construct two SDBMs, A and B, each associated
with one of the two cases: $x_i(t)=0$ and $x_i(t)=1$, respectively. The
link weights $w_{ij}^\mathrm{A}$ (or $w_{ij}^\mathrm{B}$) and the nodal
bias $b_i^\mathrm{A}$ (or $b_i^\mathrm{B}$) for node $i$ in SDBM A (or B)
are computed for all the time steps $t$ satisfying $x_i(t)=0$
(or $x_i(t)=1$), leading to
$P\{x_i(t+1)=1| \mathbf{X}^{\mathrm{R}}_i(t)\}$ for $x_i(t)=0$ (or $x_i(t)=1$)
from SDBM A (or B). Using this strategy, the dominant peaks at zero
persist in the error distributions for the dynamical processes in
category IV, as shown in Figs.~\ref{fig:all_error_distr}(g-i). For
the epidemic spreading dynamics (CP and SIS) in category V, the fixed
value of $P_i^{1\rightarrow0}$ for each node $i$ can be acquired through
$P_i^{1\rightarrow0} \simeq \langle x_i(t_1+1)\rangle_{t_1}$, where
$\langle x_i(t_1+1)\rangle_{t_1}$ stands for the average of $x_i(t_1+1)$
over all values of $t_1$ satisfying $x_i(t_1)=1$. Through this approach, SDBM
B is in fact a network without links but with each node's bias satisfying
$\mu_i = \exp(b_i)/[1+\exp(b_i)]$, where $\mu_i$ is node $i$'s recovery
rate. Figures~\ref{fig:all_error_distr}(j) and \ref{fig:all_error_distr}(k)
show the error distributions of the conditional probability recovery for
the spreading processes, where we see that the errors are essentially zero.
If the method of solving SDBM B in category IV is adopted to the dynamical
processes in other categories, i.e., without any prior knowledge about
$P_i^{1\rightarrow0}$, the resulted SDBM would have nearly identical
link weight values with respect to SDBM A (in categories I-III) or have
close-to-zero link weights and $\mu_i \simeq \exp(b_i)/[1+\exp(b_i)]$ for
category V, despite that their conditional probability recovery errors
may be slightly larger than those in Fig.~\ref{fig:all_error_distr}.
The persistent occurrence of a dominant peak at zero in the error distribution
suggests the power of combined SDBMs as a universal dynamics
approximator, regardless of the specifics of the transition
probability. When limited prior knowledge about $P_i^{1\rightarrow0}$
is available, SDBM B can be simplified or even removed without
compromising the estimation accuracy. In general, the universal
approximator has a significant short-term predictive power for arbitrary
types of dynamics on complex networks.

In the terminology of machine learning, the conditional probability
recovery error is called the ``training error,'' since it is obtained
from the same data set used to build (or ``train'') the approximator.
In machine learning, the data points generated
from the same system that have not been used in the training process can be
exploited to validate or test the actual performance of the trained
model~\cite{bishop:2006book}, which in our case is the approximator.
As a result, the time series generated from the original complex network
system after the approximator is built can be used as the test data set.
(Absence of hyper-parameters in the reconstruction process means that
cross validation is unnecessary.) In most cases,
the training errors are generally  smaller than the test errors, since the
training data set is already well-fitted by the model (the approximator)
in the training process, while the test data are new and may be out of
the fitting range of the current model. Feeding the state configurations
of the test data set into the approximator, we calculate the corresponding
conditional probabilities using Eq.~\eqref{eq:CP2} and compare them to the
true values. The results reveal a clear advantage of the approximator
built from our scheme, i.e., the training and test errors are nearly
identical, indicating the absence of any over-fitting issues that are common 
in various machine learning methodologies~\cite{bishop:2006book}.

\begin{figure}
\includegraphics[width=\linewidth]{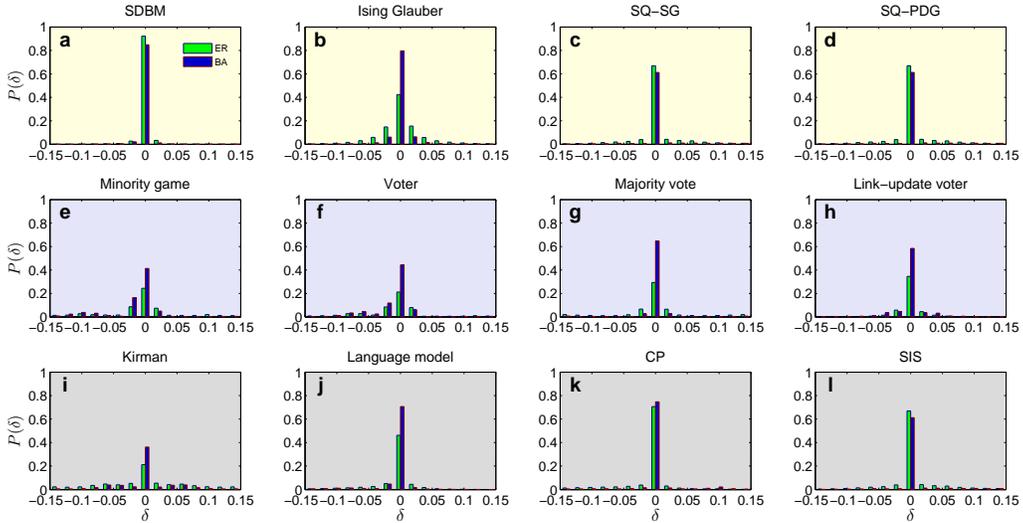}
\caption{\textbf{Distributions of the conditional probability estimation
errors for various complex network dynamics.} The distributions of
$\delta = |P\{x_i(t+1)=1| \mathbf{X}_i^\mathrm{R}(t)\}
- P^\mathrm{est}[x_i(t+1)=1| \mathbf{X}_i^\mathrm{R}(t)]|$ for the
dynamical processes in categories I to V from Table~\ref{tab:topo1} for
ER random (green bars) and BA scale-free (blue bars) networks, where
$P^\mathrm{est}[x_i(t+1)=1| \mathbf{X}_i^\mathrm{R}(t)]$ denotes the
conditional probability estimated from the corresponding SDBM approximator
through the original state configuration times series, where $1\leq i \leq N$
and $1\leq t \leq T$.}
\label{fig:all_error_distr}
\end{figure}

\section{Discussion}

Reconstructing complex dynamical systems from data is a frontier field
in network science and engineering with significant applications.
Our study was motivated by the question: is it possible to
build a universal ``machine'' to reconstruct, from data only, the
underlying complex networked dynamical system and to make predictions?
While this paper does not provide a mathematically rigorous solution, 
significant and (in some cases) striking results are obtained, which give 
strong credence that such a machine may be possible. In particular,
we combine compressive sensing and machine learning to propose two
concepts: universal network structural estimator and dynamics approximator.
Universality is fundamentally possible due to the fact that many dynamical
processes on complex networks are of the Markov type and the interactions among
the nodes are local. As a result, utilizing basic tools from machine learning 
and statistical physics, one can build up an energy based Markov network
model (e.g., a sparse dynamical Boltzmann machine) to construct a
universal estimator and approximator for different types of complex
network structures and dynamics. For a large number of representative
dynamical processes studied in this paper, we demonstrate that such
a universal SDBM can be reconstructed based on compressive
sensing and the scheme of K-means using data only, without requiring
any extra information about the network structure or the dynamical
process. The working of the SDBM is demonstrated using a large number of
combinations of the network structure and dynamics, including many
real world networks and classic evolutionary game dynamics.
An SDBM with its parameters given by the
equations constructed from the times series along with the estimated
network structure is able to reproduce the conditional probabilities
quantitatively and, accordingly, it is capable of predicting the
state configuration at least in a short term. We demonstrate that, for
certain types of dynamics, the approximator can reproduce the dynamical
process statistically, indicating the potential of its serving as a
generative model for long term prediction in such cases 
(Supplementary Note 5). 
 
While we assume binary dynamics, in principle the methodology can be
applied to other types of dynamical processes. In particular, 
Eq.~\eqref{eq:CP} can be readily extended to the conditional probability 
of each possible state value $\lambda_j$, i.e.,
$P\{x_i(t+1)=\lambda_j | \mathbf{X}^{\mathrm{R}}_i(t)\} =
P\{x_i(t+1)=\lambda_j,\mathbf{X}^{\mathrm{R}}_i(t)\}/\sum_s P\{x_i(t+1)
=\lambda_s,\mathbf{X}^{\mathrm{R}}_i(t)\}$. A potential difficulty is that
the configuration space of the system states grows exponentially with the 
number of choices of $\lambda_j$ so that, given a finite amount of data, the 
number of time instances corresponding to each particular configuration 
decreases exponentially, leading to a significant reduction in the 
estimation precision of the conditional probability.

Our effort represents an initial attempt to realize a universal
estimator and approximator, and the performance of our method is
quite competitive in comparison with the existing reconstruction
schemes designed for specific types of dynamics, such as the
compressive sensing technique for evolutionary game and epidemic spreading
dynamics~\cite{WCHLH:2009,RWLL:2010,WLGY:2011,SWFDL:2014,HSWD:2015}, as
well as methods adopting the Bethe approximation, pseudo-likelihood,
mean-field theory, and decimation operation for static and kinetic
inverse Ising problems~\cite{T:1998,CM:2011,AE:2012,NB:2012,R:2012,
RH:2011,MJ:2011,SBD:2011,Z:2012,ZAAHR:2013,BLHSB:2013,DR:2014}. In
realistic applications, the data obtained may be discontinuous or
incomplete. In such cases, the short-term predictive power possessed
by the universal estimator and approximator can be exploited to overcome
the difficulty of missing data, as the Markov network nature of the
SDBM makes backward inference possible so that the system configurations
during the time periods of missing data may be inferred. When long
term prediction is possible, the universal approximator has the
critical capability of simulating the system behavior and predicting
the chance that the system state enters into a global absorption phase,
which may find significant applications such as disaster early warning.
Another interesting reverse-engineering problem lies in the mapping
between the original dynamics and the corresponding parameter value
distribution of the reconstructed SDBM. That is, a certain parameter
distribution of the SDBM may indicate a specific type of the original
dynamics. As such, the correspondence can be used for precisely
identifying nonlinear and complex networked dynamical systems.
It is also possible to assess the relative importance of the nodes
and links in a complex network based on their corresponding biases and
weights in the reconstructed SDBM for controlling the network dynamics.
These advantages justify our idea of developing a universal machine
for data-based reverse engineering of complex networked dynamical
systems, calling for future efforts in this emerging direction to
further develop and perfect the universal network structural estimator
and dynamics approximator.

\section{Methods}

\paragraph*{Structural estimation details.}
For each type of dynamics, $10$ different network realizations are
generated for both BA scale-free and ER random topologies. For any
node in a particular network realization, the corresponding connection
weight vector $\mathbf{V}_{N\times1}$ is obtained by averaging over
$100$ compressive sensing implementations. The elements corresponding
to the existent nodes are distinguished from the non-existent links
by feeding the averaged $\mathbf{V}_{N\times1}$ into the K-means algorithm
(see below). The unweighted adjacency matrix is then obtained.

\begin{figure}
\centering
\includegraphics[width=\linewidth]{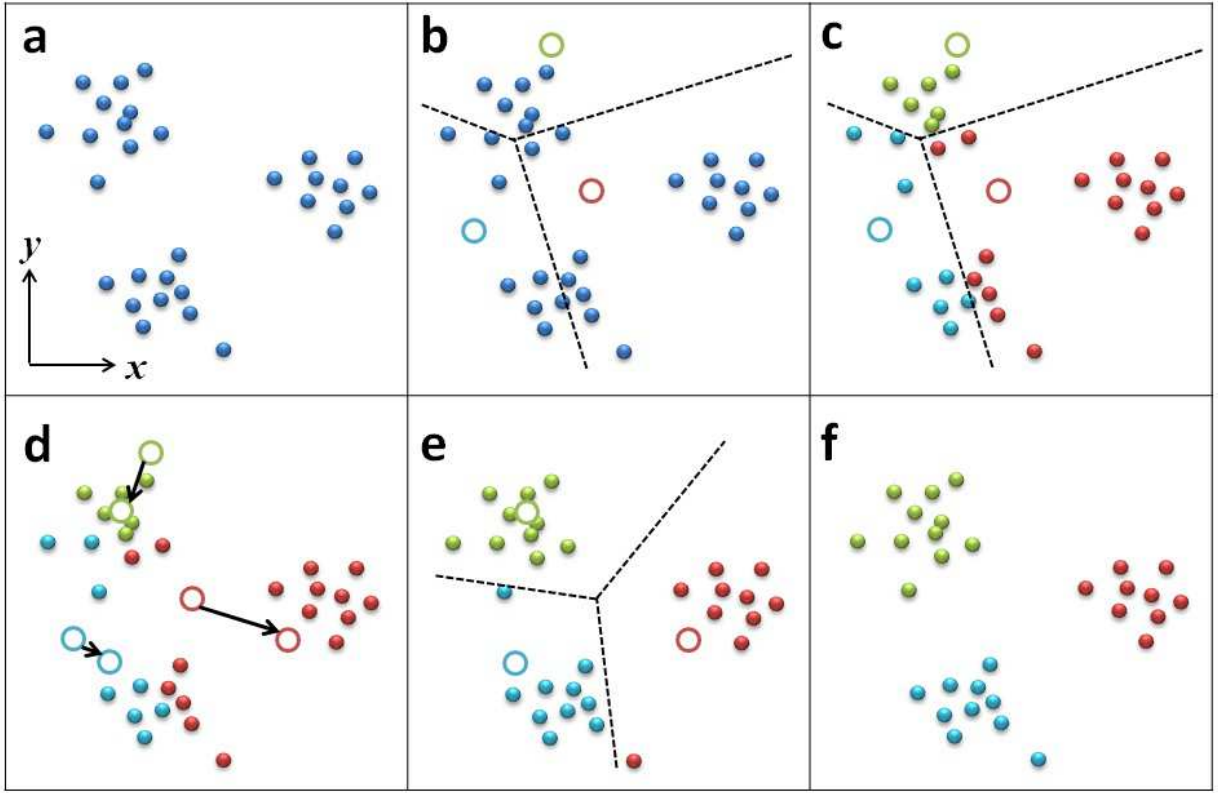}
\caption{\textbf{A schematic illustration of the K-means algorithm for 
two-dimensional data clustering}. (a) The data points (solid blue circles)
to be clustered in a 2D feature space. There are three clusters, so $K=3$. 
The K-means algorithm is capable of assigning each data point into one of 
the clusters. (b) For random locations of the cluster centers (aqua, green, 
and red hollow circles), each data point can be associated with the closest
center. (c) The 2D space is divided into three regions through three 
decision boundaries (black dashed lines), each containing the corresponding 
data points whose closest center is within. The data points currently in 
the regions with the aqua, green, and red centers are 
assigned to the corresponding clusters, respectively. At this stage, 
data points may be assigned to a cluster that did not belong to the one 
at the beginning of the process. (d) Each center moves to the centroid of 
the data points currently assigned to it (movements shown by the black arrows).
(e) The updated cluster assignments of the data points are obtained 
according to the new center locations. The steps in (c) and (d) are 
repeated until convergence is achieved. (f) The final cluster assignments. 
Different from this illustration, the compressive sensing solutions of the 
link weights $w_{ij}$ are one-dimensional data points, which form two 
clusters in 1D, corresponding to the existent and non-existent links, 
respectively. (For better visualization, we use the 2D case to illustrate 
the K-means algorithm  instead of 1D.)}
\label{fig:kmeans}
\end{figure}

\emph{Conflict resolution.} Since the links are bidirectional, the node on 
each side provides a weighted solution of the same link. The two solutions 
may be quite different, giving contradictory indication of the existence of
the link and resulting in an asymmetric adjacency matrix. For
majority types of dynamical processes on complex networks, compressive
sensing almost always gives higher prediction accuracy for lower
degree nodes due to their connection sparsity, which holds true
especially for nodes with their degrees smaller than the network
average. Based on this fact, when encountering contradictory solutions,
we determine the link existence by examining the lower degree side if
the degree value is equal to or smaller than the network average. For
cases where the degrees of both sides are larger than the network
average, we find that the false negative rates are often high. This
is because compressive sensing tends to give over simplified results
such as inducing a more than necessary level of sparsity to the solution or
causing large fluctuations with the non-existent links so they are
mixed up with the existent ones. As a result, contradictory
solutions are treated as positive (existent) solutions. For two
types of dynamics, the SDBM dynamics and link-update voter dynamics,
or dynamics on real world networks, treating all contradictory
solutions this way, regardless of degree values, can improve the
reconstruction precision, albeit only slightly in some cases.

\begin{figure}
\centering
\includegraphics[width=\linewidth]{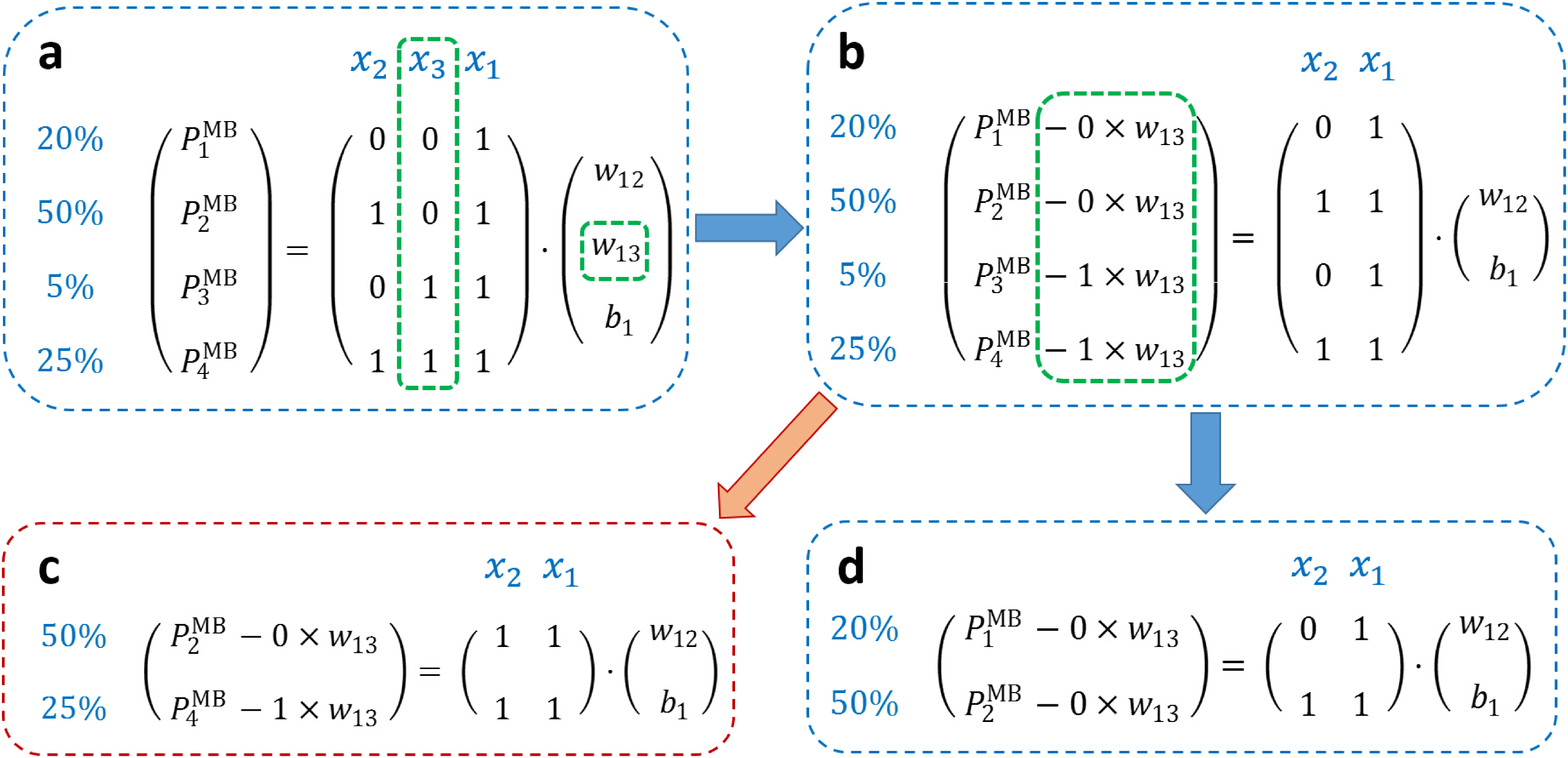}
\caption{\textbf{An example of our proposed degree guided 
solution-substitution operation and final equation group construction.}
The target node $1$ has only two neighbors, nodes $2$ and $3$. Without
loss of generality, we set $k_3<k_1=2<k_2$ so that the link weight
between nodes $1$ and $3$, $w_{13}$, is already obtained. The quantity
$P^{\mathrm{MB}}_{s} = \ln(1/\langle x_1(t+1)\rangle_s - 1)$ ($s=1,\ldots, 4$, 
as there are $2^{k_1} =4$ possible $(x_2, x_3)$ configurations
in total) denotes the estimated probability for $x_1=1$ under
configuration $s$ at the next time step. The occurring probabilities of the
corresponding $(x_2, x_3)$ configurations are listed as percentages in
the left column of each panel. (\textbf{a}) All four $(x_2, x_3)$
configurations are shown by the first two columns of the $4\times 3$
matrix in the equation. (\textbf{b}) The contribution of the known
$w_{13}$ and the configurations of $x_3$ as marked in (\textbf{a}) are
moved to the left-hand side of the equation. (\textbf{c}) In order to
solve for $(w_{12}, b_1)^{\mathrm{T}}$, the two most precise
$P_s^{\mathrm{MB}}$ values ($s=2$ and $4$) corresponding to the most
frequent $(x_2, x_3)$ configurations from (\textbf{b}) are chosen to
build a linear 2-equation group. However, the $2\times2$ matrix on
the right-hand side does not have full rank and, hence, this equation
group is ill defined (\textbf{d}) Configurations $s=1$ and $2$ are
selected to construct a 2-equation group with a full rank $2\times2$
matrix on the right-hand side and relatively precise probability
estimations, and this equation group is used to solve
$(w_{12}, b_1)^{\mathrm{T}}$.}
\label{fig:substutution}
\end{figure}

\paragraph*{Implementation of the K-means algorithm.}
K-means is one of the simplest and most popular clustering algorithms,
which has been used widely for unsupervised clustering 
tasks~\cite{bishop:2006book}. It provides each data example an 
assignment to a cluster within which the data examples are relatively 
close (or similar) while being distant from the ones in other clusters. 
The main steps for solving a typical two-dimensional data clustering 
problem via K-means are schematically illustrated in Fig.~\ref{fig:kmeans}.
For each vector $\mathbf{V}_{N\times1}$ obtained from averaging
over multiple applications of compressive sensing, picking out
the elements corresponding to the existent links from the non-existent
ones with their fluctuating element values is equivalent to a
one-dimensional clustering problems with two target clusters,
which is one dimension lower than the case shown in Fig.~\ref{fig:kmeans}. 
When implementing the K-means algorithm, we choose the initial cluster
center positions for the two clusters as the maximum and minimum
values of the vector elements with the bias $b_i$ excluded. This
is justified because the compressive sensing solution of $b_i$
can have an overwhelmingly high absolute value, which does not
provide any structural information but can severely disrupt the
clustering process, as demonstrated in Figs.~\ref{fig:solution}(d,f). Due
to the sparsity of complex networks, the cluster with smaller number
of components is regarded as containing the solutions of the
existent links, and the components of the other one correspond to
the non-existent links.

\paragraph*{Link weights and nodal bias estimation.}
For each node $i$, we rank the occurrences of all the existing
configurations of its neighbor's states. Among the top ones,
$k_i+1$ are selected to ensure that the coefficient matrix
$\mathcal{\Psi}^\mathrm{MB}_{(k_i+1)\times(k_i+1)}$ on the right
hand side of Eq.~\eqref{eq:final_Eq2} has full rank so that
the solutions, $w_{im_1}, w_{im_{2}}, \ldots, w_{im_{k_i}}, b_i$,
are unique (the selected configurations are not necessarily on
the exact top, since the real top $k_i+1$ ones may not provide a
full rank coefficient matrix).

Due to insufficient number of samples or some particular features
of a specific dynamical process, $\langle x_i(t + 1)\rangle$, the statistical
estimation of $P\{x_i(t+1)|\mathbf{X}^{\mathrm{MB}}_i(t)\}$
(or $P\{x_i(t+1)|\mathbf{X}^{\mathrm{R}}_i(t)\}$), can be $0$ or $1$,
which respectively makes $1/\langle x_i(t + 1)\rangle$ or
$\ln[1-1/\langle x_i(t + 1)\rangle]$ in Eqs.~\eqref{eq:final_Eq}
and \eqref{eq:final_Eq2} diverge. Without loss of generality, we
set $\langle x_i(t + 1)\rangle=\epsilon$ (or
$\langle x_i(t + 1)\rangle=1-\epsilon$) if $\langle x_i(t + 1)\rangle=0$
(or $1$), where $\epsilon$ is a small positive constant.

For node $i$ of degree $k_i$, the total number of the possible
neighboring state configurations is $2^{k_i}$, so large degree nodes
have significantly more configurations. The process may then regard
every particular type of configurations as useful and accordingly lead to
inaccurate estimate of $P\{x_i(t+1)|\mathbf{X}^{\mathrm{MB}}_i(t)\}$.
Our computations show that properly setting a tolerance in the Hamming
distance that allows similar but not exactly identical configurations to be
treated as the same $\mathbf{X}^{\mathrm{MB}}_i(t)$ can improve
the estimation performance. For a particular neighboring state
configuration with too fewer identical matches in the observed data,
configurations with difference in one or more digits in the Hamming
distance are used instead, until a sufficient number of matches are
found. For nodes of degrees larger than, say $15$, the Hamming
distance tolerances within $2$ or $3$ usually lead to a reasonable
number of matches for estimating $P\{x_i(t+1)|\mathbf{X}^{\mathrm{MB}}_i(t)\}$.
Extensive calculation shows that the small inaccuracy has little
effect on the reconstruction accuracy.

\paragraph*{Degree guided solution-substitution operation.}
For node $i$'s neighbors, the occurrence frequency of a particular
state configuration is generally higher for smaller values of
$k_i$. Accordingly, the probability estimation conditioned on this
configuration can be more accurate for smaller values of $k_i$.
More precise conditional probabilities in turn lead to a higher
estimation accuracy of link weights and node biases. Similar to
resolving the solution contradiction in the structural reconstruction
step, for a pair of linked nodes, $i$ and $j$ of different degrees,
$w_{ij}$ solved from the equation group of node $i$ is likely
to be more accurate than $w_{ji}$ obtained via node $j$, if $k_i < k_j$,
even ideally they should have the identical values. As a result,
using the solution obtained from a lower degree node as the value of
the link weight provides a better estimation. We run the calculation for
all nodes in the network one by one in a degree-increasing order so that
the link weights of the smallest degree nodes are acquired earlier than
for other nodes. For the larger degree nodes, the link weights shared with
the smaller degree nodes are substituted by the previously obtained
values, which are treated as known constants instead of variables
waiting to be solved. This operation effectively removes the contribution
of the lower degree neighbors from the equation and so reduces the
number of unknown variables. A reduced equation group in a form similar
to Eq.~\eqref{eq:final_Eq2} can be built up based on the most frequently
occurring state configurations of the remaining neighbors. When the full-rank
condition is met, maximum possible precision of the remaining unknown
link weights can be achieved. Consequently, with the substitution
operation, the $k_i$ link weights of node $i$, which can be inaccurate
if solved from the original $k_i+1$-equation group, can be estimated with
the maximum possible accuracy. An example of this substitution operation
and the final equation group construction process is shown in
Fig.~\ref{fig:substutution}.

Our calculations show that the application of the degree guided
solution-substitution operation can significantly increase the
accuracy of the estimation of the link weights and nodes biases. This
operation is thus key to making the SDBM correspondence
a reliable approximator of the original network dynamics.

As shown in Fig.~\ref{fig:all_error_distr}, for dynamical processes
in categories I and II, the estimation errors are dominantly distributed
at zero. For processes in other categories, besides the dominant peak
at zero, there exist multiple small peaks at nonzero error values. These
small side peaks are a consequence of the model complexity of the Markov
networks.

\paragraph*{Model complexity and representation power.}
For dynamical processes in categories I and II, the transition
probabilities $P_i^{0\rightarrow1}$ all have the form
$1/[1+\exp(A n_i + B k_i)]$, where $A$ and $B$ are constants, and $n_i$
denotes the number of node $i$'s neighbors whose states are $1$. We have
\begin{displaymath}
[1+\exp(A n_i + B k_i)]^{-1} =
\{1+\exp[\sum_{m=1}^{k_i} w_{im} x_m(t) + b_i]\}^{-1},
\end{displaymath}
which gives
\begin{equation} \label{eq:linear}
\sum_{m=1}^{k_i} A x_m(t) + B k_i = \sum_{m=1}^{k_i} w_{im} x_m(t) + b_i.
\end{equation}
In the ideal case where an absolutely accurate estimate of
$P\{x_i(t+1)=1|\mathbf{X}^{\mathrm{R}}_i(t)\}$ can be obtained,
Eq.~\eqref{eq:linear} holds for any possible neighboring state configurations.
We thus have $w_{im} = A$ and $b_i = B k_i$, and the probabilities
conditioned on these configurations sharing the same values of $n_i$ in
the approximator are all equal to $P_i^{0\rightarrow1}$. This means
that, in theory, the conditional probabilities can be reconstructed
with zero errors. In fact, a one-to-one mapping between the coefficients
indicates that the model complexity of the SDBM provides the approximator
with sufficient representation power to model the dynamics processes
in categories I and II. Practically, since the statistical estimations
of $P\{x_i(t+1)=1|\mathbf{X}^{\mathrm{R}}_i(t)\}$ may not be absolutely
identical under different neighboring configurations with the same values
of $n_i$ due to random fluctuations, we have $w_{im} \approx A$ so that
the conditional probabilities generated by the approximator
may differ from each other slightly and also from the true probability
$P_i^{0\rightarrow1}$. As a result, random recovery errors can occur.

For dynamical processes in categories other than I and II, the simple
coefficient-mapping relation between the corresponding $x_m(t)$ terms
on the two sides of Eq.~\eqref{eq:linear} become nonlinear, due
to the fact that the specific forms of $P_i^{0\rightarrow1}$ differ
substantially from that of the SDBM. In this case, each particular
neighboring state configuration produces a distinct equation. There
are $2^{k_i}$ equations in total, while the number of unknown variables
to be solved is only $k_i + 1$ ($w_{im}$ for $m=1,\ldots,k_i$ and $b_i$).
For $k_i\geq 2$, there are thus more equations than the number of unknown
variables. As a result, the representation power of the SDBM approximator
is limited by its finite model complexity so that, even in principle,
the approximator may not be able to fully describe the
original dynamical process. In our framework, we calculate the Markov
link weights and the node biases according to the $k_i+1$ most frequently
appeared neighboring state configurations. A consequence is that
imprecise conditional probability estimations can arise for the
configurations with relatively lower occurring frequency, giving rise
to the nonzero peaks in the error distribution in
Fig.~\ref{fig:all_error_distr}. However, interestingly, with respect
to the precision of the conditional probabilities produced by the
approximator, the SDBM parameters do not show a significantly strong
preference towards the most frequently occurring configurations. In
practice, the estimated conditional probabilities for the majority of
the less frequently occurring configurations also fall within a close
range of the true values. This phenomenon suggests the power of the
approximator to work beyond the limit set by its theoretical model
complexity. This is the main reason for the emergence and persistence
of the dominant peak at zero in the error distribution.


\section*{Acknowledgements}

This work was supported by ARO under Grant No.~W911NF-14-1-0504
and by NSF under Grant No.~1441352.

\section*{Author contributions}

Y.-Z.C. and Y.-C.L. designed research; Y.-Z.C. performed research;
both analyzed data; Y.-C.L. and Y.-Z.C. wrote the paper.

\section*{Additional information}

Competing financial interests: The authors declare no competing
financial interests.

\end{document}